\documentclass[11pt,showpacs,showkeys,preprintnumbers,amsmath,amssymb]{revtex4}

\begin{document}


\title{Historical pseudo simplified solution of the Dirac-Coulomb equation}
\author{Ruida Chen³Âï}

\affiliation{Shenzhen Institute of Mathematics and Physics,
Shenzhen, 518028, China}


\begin{abstract}

One of the simplified solutions of the Dirac equations with the pure
Coulomb potential given in a paper published in 1985 is pseudo. The
original paper solved the Dirac equations by introducing a
transformation of functions with two strange parameters $a$ and $b$
to transform the original system of the first-order differential
equations into two uncoupled differential equations of second order.
However, not only the given eigenvalues sets violate the uniqueness
of solution but also the said second-order equations are not any
necessarily mathematical deduction. In order to determine the
introduced parameters, formally, the author actually introduced some
self-contradictory mathematical formulas, such as $\sinh \theta =
2ab$, $\cosh \theta = a^2 + b^2$, $\tanh \theta = - {Z\alpha }
\mathord{\left/ {\vphantom {{Z\alpha } k}} \right.
\kern-\nulldelimiterspace} k$, $a^2 - b^2 = 1$, $b = \sinh \left(
{\theta \mathord{\left/ {\vphantom {\theta 2}} \right.
\kern-\nulldelimiterspace} 2} \right)$ and $a = \cosh \left( {\theta
\mathord{\left/ {\vphantom {\theta 2}} \right.
\kern-\nulldelimiterspace} 2} \right)$. But one has not known the
value of the parameters $a$ and $b$ all the while, whereas the
parameters were insensibly deleted in the given second-order
Dirac-Coulomb equation last. One cannot recover any result given in
the paper by making corresponding correctly mathematical
calculations.

\end{abstract}

\maketitle

\section{Introduction}

It is well known that the exact solutions of the wave
equations\cite{Bialynicki:2004} are very important. Because the
judgment on whether or not the approximately solution of the wave
equations exist usually dependents on the existence of exact
solutions and some real laws\cite{Bialynicki:1997}\cite{Cooper:2003}
are often obtained from the exact solution of the wave equations.
For the Dirac equation for a single particle in the Coulomb field,
quantum mechanics textbooks generally adopt the exact solution that
was first given by Darwin and
Gordon\cite{Darwin:1928}\cite{Gordon:1928}. Of course, there are
some mathematical problems needing to discuss in the Darwin-Gordon
solution\cite{Chen:2007}.$^{ }$However, some other exact solutions
of the second-order Dirac-Coulomb equation appear simplified but are
clearly incorrect in character\cite{Chen:1}\cite{Chen:2}. In form,
it derived those so-called simplified solution by first transforming
the system of the first-order Driac-Coulomb equation into the
Schr\"{o}dinger-like or the Klein-Gordon-like equations then solving
the corresponding second-order differential equations to write the
distinguished energy eigenvalues. Nevertheless, one cannot recover
the corresponding solution basing on the introduced mathematical
methods in the original papers. It has been pointed out that many
given solutions and the corresponding formulas of the energy levels
are actually not the necessary mathematical deductions of the said
second-order Dirac-Coulomb equation. In addition, it is worse that
some so-called second-order Dirac-Coulomb equations given in those
published papers can not be yielded from the original system of the
Dirac- equations of first-order in the Coulomb field. They are not
any necessary mathematical deduction of physics and mathematics yet.

Here we show that a historically formal simplified solution of the
Dirac-Coulomb equation given in a paper\cite{Simplified:1985}
published in Physical Review 22 years ago is a pseudo solution. For
the said second-order Dirac-Coulomb equations in which two equations
were written in the same form by using sign ``$\pm $'', two
eigenvalues set should be given and they are actually different. It
is well known that two different sets of the energy eigenvalues for
the same quantum system violate the uniqueness of solution. However,
in the original paper, only one of the eigenvalues sets was given
and the other was thrown out of all reason. This case still exists
in other papers
today\cite{Hakan:2005}\cite{Nicolas:2006}\cite{Alhaidari:2007}. In
particular, in order to write the second-order Dirac-Coulomb
equation, the author introduced two strange parameters $a$ and $b$
and afterward gave some self-contradictory expressions such as
$\sinh \theta = 2ab$, $\cosh \theta = a^2 + b^2$, $\tanh \theta = -
{Z\alpha } \mathord{\left/ {\vphantom {{Z\alpha } k}} \right.
\kern-\nulldelimiterspace} k$, $a^2 - b^2 = 1$, $b = \sinh \left(
{\theta \mathord{\left/ {\vphantom {\theta 2}} \right.
\kern-\nulldelimiterspace} 2} \right)$ and $a = \cosh \left( {\theta
\mathord{\left/ {\vphantom {\theta 2}} \right.
\kern-\nulldelimiterspace} 2} \right)$. However, one has not known
the values of parameter$a$ and $b$. By using the mathematical method
introduced therein or by using other correct mathematical methods,
one cannot recover any so-called Schr\"{o}dinger-like or
Klein-Gordon-like equations given by the author. Consequently, in
the mentioned paper, the claimed simplified solutions of the
Dirac-Coulomb equation, the corresponding formula for the energy
levels and the corresponding mathematical procedures are pseudo.

\section{Original formal simplified solution of Dirac-Coulomb equation}

Many authors claimed that they obtain simplified solution of the
Driac-Coulomb equation. In 1985, Su considered that, with the use of
a simple similarity transformation which brought the radial wave
equations of the Dirac-Coulomb problem into a form nearly identical
to those of the Schr\"{o}dinger and Klein-Gordon equations, he
derived simplified solutions to the Dirac-Coulomb equation for both
the bound and continuum states following the familiar standard
procedurfe adopted in the derivation of the conventional solutions.
He considered that to obtain the desired form of the second-order
radial equations he could still work with a first-order partial
differential equation rather than with the second-order Dirac
equation widely employed in the derivation of the simplified
solutions, and thus he can avoid the task of reducing the solutions
of the second-order equations to those of the original Dirac
equations. Here we only check the mathematical procedure for
deducing the eneregy eigensolutions for the bound state in the
original paper. The author first wrote the radial Dirac-Coulomb
equation in the following form
\begin{equation}
\label{eq1} {H}'_r \left( {{\begin{array}{*{20}c}
 {R\left( r \right)} \hfill \\
 {Q\left( r \right)} \hfill \\
\end{array} }} \right) = E\left( {{\begin{array}{*{20}c}
 {R\left( r \right)} \hfill \\
 {Q\left( r \right)} \hfill \\
\end{array} }} \right)
\end{equation}

\noindent with
\begin{equation}
\label{eq2} {H}'_r = \left( {{\begin{array}{*{20}c}
 {A}' \hfill & {B}' \hfill \\
 {C}' \hfill & {D}' \hfill \\
\end{array} }} \right)
\end{equation}

\noindent where
\begin{equation}
\label{eq3}
\begin{array}{l}
 {A}' = mc^2\cosh \theta + \hbar c\left[ {\sinh \theta \left( {\frac{d}{dr} + \frac{1}{r}} \right) - \frac{Z\alpha }{r}} \right] \\
 {B}' = - \left\{ {mc^2\sinh \theta + \hbar c\left[ {\cosh \theta \left( {\frac{d}{dr} + \frac{1}{r}} \right) - \frac{k}{r}} \right]} \right\} \\
 {C}' = mc^2\sinh \theta + \hbar c\left[ {\cosh \theta \left( {\frac{d}{dr} + \frac{1}{r}} \right) + \frac{k}{r}} \right] \\
 {D}' = - \left\{ {mc^2\cosh \theta + \hbar c\left[ {\sinh \theta \left( {\frac{d}{dr} + \frac{1}{r}} \right) + \frac{Z\alpha }{r}} \right]} \right\} \\
 \end{array}
\end{equation}

\noindent and $k = \pm \left( {j + \frac{1}{2}} \right)$, $\alpha =
\frac{e^2}{\hbar c}$ being the fine-structure constant. The strange
parameters $a$ and $b$ were claimed the real constants by the
author. Then it was introduced that
\begin{equation}
\label{eq4} \cosh \theta = \frac{a^2 + b^2}{a^2 - b^2},\quad \sinh
\theta = \frac{2ab}{a^2 - b^2},\quad a^2 - b^2 > 0
\end{equation}

\noindent and it was selected that
\begin{equation}
\label{eq5} a^2 - b^2 = 1,\quad a = \cosh \left( {\frac{\theta }{2}}
\right),\quad b = \sinh \left( {\frac{\theta }{2}} \right)
\end{equation}

\noindent with At the same time, the author also introduced the
other function
\begin{equation}
\label{eq6} h\theta = { - Z\alpha } \mathord{\left/ {\vphantom {{ -
Z\alpha } K}} \right. \kern-\nulldelimiterspace} K
\end{equation}

\noindent It was alleged that one attained great simplification in
solving the radial equations
\begin{equation}
\label{eq7}
\begin{array}{l}
 Q\left( r \right) = \left[ { - \frac{E\tilde {\omega }Z\alpha }{\gamma } + \hbar c\left( {\frac{d}{dr} + \frac{1 + \tilde {\omega }\gamma }{r}} \right)} \right]\frac{R\left( r \right)}{mc^2 + \left( {j + \frac{1}{2}} \right)\frac{E}{\gamma }} \\
 R\left( r \right) = \left[ {\frac{E\tilde {\omega }Z\alpha }{\gamma } + \hbar c\left( {\frac{d}{dr} + \frac{1 - \tilde {\omega }\gamma }{r}} \right)} \right]\frac{Q\left( r \right)}{mc^2 - \left( {j + \frac{1}{2}} \right)\frac{E}{\gamma }} \\
 \end{array}
\end{equation}

\noindent where $\tilde {\omega } = \mp 1$, $\gamma = \left[ {\left(
{j + \frac{1}{2}} \right)^2 - Z^2\alpha ^2} \right]^{1
\mathord{\left/ {\vphantom {1 2}} \right. \kern-\nulldelimiterspace}
2}$. Finally, without some mathematical calculations, it was also
alleged that one should derive the equation
\begin{equation}
\label{eq8} \left[ {\frac{1}{r^2}\frac{d}{dr}\left(
{r^2\frac{d}{dr}} \right) + \left( {\frac{E^2 - m^2c^4}{\hbar ^2c^2}
+ \frac{2EZ\alpha }{\hbar cr}} \right) - \frac{\gamma ^2\pm \tilde
{\omega }\gamma }{r^2}} \right]\times \left( {{\begin{array}{*{20}c}
 {R\left( r \right)} \hfill \\
 {Q\left( r \right)} \hfill \\
\end{array} }} \right) = 0
\end{equation}

\noindent and the distinguished Dirac formula of the energy levels
in the Coulomb field for bound state
\begin{equation}
\label{eq9} E \mathord{\left/ {\vphantom {E {mc^2}}} \right.
\kern-\nulldelimiterspace} {mc^2} = \left[ {1 + {Z^2\alpha ^2}
\mathord{\left/ {\vphantom {{Z^2\alpha ^2} {\left( {n_r + \sqrt
{\left( {j + \frac{1}{2}} \right)^2 - Z^2\alpha ^2} } \right)^2}}}
\right. \kern-\nulldelimiterspace} {\left( {n_r + \sqrt {\left( {j +
\frac{1}{2}} \right)^2 - Z^2\alpha ^2} } \right)^2}} \right]^{{ - 1}
\mathord{\left/ {\vphantom {{ - 1} 2}} \right.
\kern-\nulldelimiterspace} 2}
\end{equation}

\section{Original eigensolutions sets violate the uniqueness of solution}

In fact, the system of the second-order equations is not always
equivalent to the corresponding first-order differential equations.
Firstly, we don not know what it means for the parameters $a$ and
$b$and how to eliminate the two parameters to derive the
second-order equations (\ref{eq8}) from those new definitions such
as from (\ref{eq1}) to (\ref{eq7}). We notice that the equation
(\ref{eq8}) given in the original article should include four
equations as follows
\begin{equation}
\label{eq10}
\begin{array}{l}
 \frac{1}{r^2}\frac{d}{dr}\left( {r^2\frac{dR}{dr}} \right) + \left( {\frac{E^2 - m^2c^4}{\hbar ^2c^2} + \frac{2EZ\alpha }{\hbar cr} - \frac{\gamma ^2 - \gamma }{r^2}} \right)R = 0 \\
 \frac{1}{r^2}\frac{d}{dr}\left( {r^2\frac{dR}{dr}} \right) + \left( {\frac{E^2 - m^2c^4}{\hbar ^2c^2} + \frac{2EZ\alpha }{\hbar cr} - \frac{\gamma ^2 + \gamma }{r^2}} \right)R = 0 \\
 \frac{1}{r^2}\frac{d}{dr}\left( {r^2\frac{dQ}{dr}} \right) + \left( {\frac{E^2 - m^2c^4}{\hbar ^2c^2} + \frac{2EZ\alpha }{\hbar cr} - \frac{\gamma ^2 - \gamma }{r^2}} \right)Q = 0 \\
 \frac{1}{r^2}\frac{d}{dr}\left( {r^2\frac{dQ}{dr}} \right) + \left( {\frac{E^2 - m^2c^4}{\hbar ^2c^2} + \frac{2EZ\alpha }{\hbar cr} - \frac{\gamma ^2 + \gamma }{r^2}} \right)Q = 0 \\
 \end{array}
\end{equation}

\noindent Each of the equations has its own eigenvalus and
eigensolutions set, and these eigenvalues are usually different from
each other. It is incorrect for giving only one of the formulas of
the energy levels and alleging to recover the distinguished Dirac
formula in mathematical and physical logic. One easily finds the
eigenvalue sect of the general equations
\begin{equation}
\label{eq11} \frac{1}{r^2}\frac{d}{dr}\left( {r^2\frac{d\Lambda
}{dr}} \right) + \left( {\frac{E^2 - m^2c^4}{\hbar ^2c^2} +
\frac{2EZ\alpha }{\hbar cr} - \frac{\gamma ^2 \mp \gamma }{r^2}}
\right)\Lambda = 0
\end{equation}

\noindent Because some details are uselly ignored by using the
corresponding special function, we directly solve every differential
equation to find the energy eigenvalues. By introducing the
substitution
\begin{equation}
\label{eq12} \Lambda = \frac{M}{r}
\end{equation}

\noindent The equation (\ref{eq11}) becomes
\begin{equation}
\label{eq13} \frac{d^2M}{dr^2} + \left( { - \frac{m^2c^4 -
E^2}{\hbar ^2c^2} + \frac{2Z\alpha }{\hbar cr}E - \frac{\gamma ^2
\mp \gamma }{r^2}} \right)M = 0
\end{equation}

\noindent This equation has the asymptotic solutions with $E < mc^2$
satisfying the boundary condition
\begin{equation}
\label{eq14} \quad M = \exp \left( { - \sqrt {\frac{m^2c^4 -
E^2}{\hbar ^2c^2}} r} \right)\quad \left( {r \to \infty } \right)
\end{equation}

\noindent It is assumed that the formal solution of the equation
(\ref{eq13}) takes the form
\begin{equation}
\label{eq15} M = \exp \left( { - \sqrt {\frac{m^2c^4 - E^2}{\hbar
^2c^2}} r} \right)u
\end{equation}

\noindent We have
\begin{equation}
\label{eq16}
\begin{array}{l}
 \frac{dM}{dr} = \exp \left( { - \sqrt {\frac{m^2c^4 - E^2}{\hbar ^2c^2}} r} \right)\left( {\frac{du}{dr} - \sqrt {\frac{m^2c^4 - E^2}{\hbar ^2c^2}} u} \right) \\
 \frac{d^2M}{dr^2} = \exp \left( { - \sqrt {\frac{m^2c^4 - E^2}{\hbar ^2c^2}} r} \right)\left( {\frac{d^2u}{dr^2} - 2\sqrt {\frac{m^2c^4 - E^2}{\hbar ^2c^2}} \frac{du}{dr} + \frac{m^2c^4 - E^2}{\hbar ^2c^2}u} \right) \\
 \end{array}
\end{equation}

\noindent Substituting into the equation (\ref{eq13}), we obtain
\begin{equation}
\label{eq17} \frac{d^2u}{dr^2} - 2\sqrt {\frac{m^2c^4 - E^2}{\hbar
^2c^2}} \frac{du}{dr} + \left( {\frac{2EZ\alpha }{\hbar cr} -
\frac{\gamma ^2 \mp \gamma }{r^2}} \right)u = 0
\end{equation}

\noindent Now, seeking the power series solution, let
\begin{equation}
\label{eq18} u = \sum\limits_{n = 0}^\infty {d_n } r^{s + n}
\end{equation}

\noindent Hence
\begin{equation}
\label{eq19} \frac{du}{dr} = \sum\limits_{n = 0}^\infty {\left( {s +
n} \right)d_n } r^{s + n - 1},\quad \frac{d^2u}{dr^2} =
\sum\limits_{n = 0}^\infty {\left( {s + n} \right)\left( {s + n - 1}
\right)d_n } r^{s + n - 2}
\end{equation}

\noindent Substitute (\ref{eq18}) and (\ref{eq19}) into the
equations (\ref{eq17}), we have
\begin{equation}
\label{eq20} \sum\limits_{n = 0}^\infty {\left\{ {\begin{array}{l}
 \left[ {\left( {s + n} \right)\left( {s + n - 1} \right) - \left( {\gamma ^2 \mp \gamma } \right)} \right]d_n \\
 - \left[ {2\sqrt {\frac{m^2c^4 - E^2}{\hbar ^2c^2}} \left( {s + n - 1} \right) - \frac{2EZ\alpha }{\hbar c}} \right]d_{n - 1} \\
 \end{array}} \right\}r^{s + n - 2}} = 0
\end{equation}

\noindent It gives the recursive relation of the coefficients of the
power series
\begin{equation}
\label{eq21} \left[ {\left( {s + n} \right)\left( {s + n - 1}
\right) - \left( {\gamma ^2 \mp \gamma } \right)} \right]d_n -
\left[ {2\sqrt {\frac{m^2c^4 - E^2}{\hbar ^2c^2}} \left( {s + n - 1}
\right) - \frac{2EZ\alpha }{\hbar c}} \right]d_{n - 1} = 0
\end{equation}

\noindent The power series (\ref{eq18}) naturally give the initial
value condition: $d_{ - 1} = d_{ - 2} = \cdots = 0$ and $d_0 \ne 0$.
Putting let $n = 0$ and substituting it into the recursive relation
(\ref{eq21}) reads $s\left( {s - 1} \right) - \left( {\gamma ^2 \mp
\gamma } \right) = 0$, it gives
\begin{equation}
\label{eq22} s = \frac{1\pm \left( {2\gamma \mp 1} \right)}{2}
\end{equation}

\noindent Thus we have the multi-values of $s$ for the four
equations in (\ref{eq10}) respectively
\begin{equation}
\label{eq23} s_{R_1 } = \left\{ {{\begin{array}{*{20}c}
 \gamma \hfill \\
 {1 - \gamma } \hfill \\
\end{array} }} \right.,\quad s_{R_2 } = \left\{ {{\begin{array}{*{20}c}
 {1 + \gamma } \hfill \\
 { - \gamma } \hfill \\
\end{array} }} \right.,\quad s_{Q_1 } = \left\{ {{\begin{array}{*{20}c}
 \gamma \hfill \\
 {1 - \gamma } \hfill \\
\end{array} }} \right.,\quad s_{Q_2 } = \left\{ {{\begin{array}{*{20}c}
 {1 + \gamma } \hfill \\
 { - \gamma } \hfill \\
\end{array} }} \right.
\end{equation}

\noindent Since the wave function has to be normalizable we must
choose the value of $s$ to be more than$1$ but not only positive
sign. In form, for the solutions (\ref{eq23}), we can but choose
\begin{equation}
\label{eq24} s_{R_1 } = \gamma ,\quad s_{R_2 } = 1 + \gamma ,\quad
s_{Q_1 } = \gamma ,\quad s_{Q_2 } = 1 + \gamma
\end{equation}

Combining the expression (\ref{eq12}), (\ref{eq15}), (\ref{eq18}),
(\ref{eq24}), and making using of the value of $\gamma $, the
completely formal solution of the equations (\ref{eq10}) can be
written as follows
\begin{equation}
\label{eq25}
\begin{array}{l}
 R_1 = \exp \left( { - \sqrt {\frac{m^2c^4 - E^2}{\hbar ^2c^2}} r} \right)\sum\limits_{n = 0}^\infty {d_n } r^{\sqrt {\left( {j + 1 \mathord{\left/ {\vphantom {1 2}} \right. \kern-\nulldelimiterspace} 2} \right)^2 - Z^2\alpha ^2} + n - 1} \\
 R_2 = \exp \left( { - \sqrt {\frac{m^2c^4 - E^2}{\hbar ^2c^2}} r} \right)\sum\limits_{n = 0}^\infty {d_n } r^{\sqrt {\left( {j + 1 \mathord{\left/ {\vphantom {1 2}} \right. \kern-\nulldelimiterspace} 2} \right)^2 - Z^2\alpha ^2} + n} \\
 Q_1 = \exp \left( { - \sqrt {\frac{m^2c^4 - E^2}{\hbar ^2c^2}} r} \right)\sum\limits_{n = 0}^\infty {d_n } r^{\sqrt {\left( {j + 1 \mathord{\left/ {\vphantom {1 2}} \right. \kern-\nulldelimiterspace} 2} \right)^2 - Z^2\alpha ^2} + n - 1} \\
 Q_2 = \exp \left( { - \sqrt {\frac{m^2c^4 - E^2}{\hbar ^2c^2}} r} \right)\sum\limits_{n = 0}^\infty {d_n } r^{\sqrt {\left( {j + 1 \mathord{\left/ {\vphantom {1 2}} \right. \kern-\nulldelimiterspace} 2} \right)^2 - Z^2\alpha ^2} + n} \\
 \end{array}
\end{equation}

\noindent When $j = 1 \mathord{\left/ {\vphantom {1 2}} \right.
\kern-\nulldelimiterspace} 2$, $\gamma = \sqrt {1 - Z^2\alpha ^2} <
1$, the first and the third expression are divergent at the origin
of the coordinate system
\begin{equation}
\label{eq26}
\begin{array}{l}
 \mathop {\lim }\limits_{r \to 0} R_1 = \mathop {\lim }\limits_{r \to 0} \exp \left( { - \sqrt {\frac{m^2c^4 - E^2}{\hbar ^2c^2}} r} \right)\sum\limits_{n = 0}^\infty {d_n } r^{\sqrt {\left( {j + 1 \mathord{\left/ {\vphantom {1 2}} \right. \kern-\nulldelimiterspace} 2} \right)^2 - Z^2\alpha ^2} + n - 1} = \infty \\
 \mathop {\lim }\limits_{r \to 0} Q_1 = \mathop {\lim }\limits_{r \to 0} \exp \left( { - \sqrt {\frac{m^2c^4 - E^2}{\hbar ^2c^2}} r} \right)\sum\limits_{n = 0}^\infty {d_n } r^{\sqrt {\left( {j + 1 \mathord{\left/ {\vphantom {1 2}} \right. \kern-\nulldelimiterspace} 2} \right)^2 - Z^2\alpha ^2} + n - 1} = \infty \\
 \end{array}
\end{equation}

\noindent implying that the first equation and the third equation in
(\ref{eq10}) have no solution which satisfy the boundary conditions.
We know that the above divergence have been called ``mild
divergence'' by someone. In (\ref{eq25}), the second and the forth
expression are finite at the origin of the coordinate system
\begin{equation}
\label{eq27}
\begin{array}{l}
 \mathop {\lim }\limits_{r \to 0} R_2 = \mathop {\lim }\limits_{r \to 0} \exp \left( { - \sqrt {\frac{m^2c^4 - E^2}{\hbar ^2c^2}} r} \right)\sum\limits_{n = 0}^\infty {d_n } r^{\sqrt {\left( {j + 1 \mathord{\left/ {\vphantom {1 2}} \right. \kern-\nulldelimiterspace} 2} \right)^2 - Z^2\alpha ^2} + n} = 0 \\
 \mathop {\lim }\limits_{r \to 0} Q_2 = \mathop {\lim }\limits_{r \to 0} \exp \left( { - \sqrt {\frac{m^2c^4 - E^2}{\hbar ^2c^2}} r} \right)\sum\limits_{n = 0}^\infty {d_n } r^{\sqrt {\left( {j + 1 \mathord{\left/ {\vphantom {1 2}} \right. \kern-\nulldelimiterspace} 2} \right)^2 - Z^2\alpha ^2} + n} = 0 \\
 \end{array}
\end{equation}

\noindent implying that the second and the forth equation in
(\ref{eq10}) seem to have the significative solutions. Comparing
(\ref{eq26}) and (\ref{eq27}) constructs the first kind of
contradiction to the second-order differential equation
(\ref{eq10}).

On the other hand, if we accept the subjective definition of the
so-called ``mild divergence'' or ``weak divergence'' we would have
two eigenvalues set corresponding to the formula of energy levels in
the Coulomb field. Form (\ref{eq25}), combining the first and the
third expression in one form denoted by $\Lambda _1 $ and combining
the second and the forth expression in another form denoted by
$\Lambda _2 $ respectively yield
\begin{equation}
\label{eq28}
\begin{array}{l}
 \Lambda _1 = \exp \left( { - \sqrt {\frac{m^2c^4 - E^2}{\hbar ^2c^2}} r} \right)\sum\limits_{n = 0}^\infty {d_n } r^{\sqrt {\left( {j + 1 \mathord{\left/ {\vphantom {1 2}} \right. \kern-\nulldelimiterspace} 2} \right)^2 - Z^2\alpha ^2} + n - 1} \\
 \Lambda _2 = \exp \left( { - \sqrt {\frac{m^2c^4 - E^2}{\hbar ^2c^2}} r} \right)\sum\limits_{n = 0}^\infty {d_n } r^{\sqrt {\left( {j + 1 \mathord{\left/ {\vphantom {1 2}} \right. \kern-\nulldelimiterspace} 2} \right)^2 - Z^2\alpha ^2} + n} \\
 \end{array}
\end{equation}

\noindent In order that the wave functions remain normalizable we
must require that the series for $u$ so the recursive relation
(\ref{eq21}) terminate at any term with the power $n_r $, that means
that $d_{n_r } \ne 0$ and $d_{n_r + 1} = d_{n_r + 2} = \cdots = 0$.
Substituting $n = n_{r + 1} $ into (\ref{eq21}), we have
\begin{equation}
\label{eq29} 2\sqrt {\frac{m^2c^4 - E^2}{\hbar ^2c^2}} \left( {s +
n_r } \right) - \frac{2EZ\alpha }{\hbar c} = 0
\end{equation}

\noindent So that we obtain the Dirac formula in form
\begin{equation}
\label{eq30} E = \frac{mc^2}{\sqrt {1 + \frac{Z^2\alpha ^2}{\left(
{n_r + s} \right)^2}} }
\end{equation}

\noindent According to (\ref{eq24}), we finally obtain the
eigenvalues of the energy
\begin{equation}
\label{eq31}
\begin{array}{l}
 E_1 = \frac{mc^2}{\sqrt {1 + \frac{Z^2\alpha ^2}{\left( {n_r + \sqrt {\left( {j + 1 \mathord{\left/ {\vphantom {1 2}} \right. \kern-\nulldelimiterspace} 2} \right)^2 - Z^2\alpha ^2} } \right)^2}} } \\
 E_2 = \frac{mc^2}{\sqrt {1 + \frac{Z^2\alpha ^2}{\left( {n_r + 1 + \sqrt {\left( {j + 1 \mathord{\left/ {\vphantom {1 2}} \right. \kern-\nulldelimiterspace} 2} \right)^2 - Z^2\alpha ^2} } \right)^2}} } \\
 \end{array}
\end{equation}

\noindent It implies that the solutions of the equations (\ref{eq8})
given in the original paper violate the uniqueness of solution of
the wave equations. This is the serious mathematical and physical
mistakes. In addition, the formula (\ref{eq9}) is not a necessarily
mathematical and physical deduction. It constructs the second kind
of contradiction to the second-order differential equation
(\ref{eq10}).

\section{The second-order Dirac equation (\ref{eq21}) is pseudo}

In fact, one cannot transform the original radial Dirac-Coulomb
equations with first-order into the Schr\"{o}dinger-like equation or
the Klein-Gordon-like equation with the second-order (\ref{eq8}). It
means that the equation is not equivalent to the original system of
first-order Dirac-Coulomb equations. Trying to explain the meaning
of the definition such as $\sinh \theta $, $\cosh \theta $ and
$\tanh \theta $ make us puzzled, the expressions (\ref{eq4}),
(\ref{eq5}) and (\ref{eq6}) read
\begin{equation}
\label{eq32}
\begin{array}{l}
 \sinh \theta = 2ab,\quad \cosh \theta = a^2 + b^2,\quad \tanh \theta = - {Z\alpha } \mathord{\left/ {\vphantom {{Z\alpha } k}} \right. \kern-\nulldelimiterspace} k \\
 \;a^2 - b^2 = 1,\quad a = \cosh \left( {\theta \mathord{\left/ {\vphantom {\theta 2}} \right. \kern-\nulldelimiterspace} 2} \right),\quad b = \sinh \left( {\theta \mathord{\left/ {\vphantom {\theta 2}} \right. \kern-\nulldelimiterspace} 2} \right) \\
 \end{array}
\end{equation}

\noindent These expressions given in the original paper early or
late are in contradiction with each other. Why did not the author
direct give the value of $a$ and $b$, but deleted them finally
expression like magic without any mathematical operation?. Now
writing the equations (\ref{eq1}) with (\ref{eq2}) and (\ref{eq3})
in the separate form as follows
\begin{equation}
\label{eq33}
\begin{array}{l}
 \left\{ {mc^2\cosh \theta + \hbar c\left[ {\sinh \theta \left( {\frac{d}{dr} + \frac{1}{r}} \right) - \frac{Z\alpha }{r}} \right]} \right\}R \\
 - \left\{ {mc^2\sinh \theta + \hbar c\left[ {\cosh \theta \left( {\frac{d}{dr} + \frac{1}{r}} \right) - \frac{k}{r}} \right]} \right\}Q = ER \\
 \left\{ {mc^2\sinh \theta + \hbar c\left[ {\cosh \theta \left( {\frac{d}{dr} + \frac{1}{r}} \right) + \frac{k}{r}} \right]} \right\}R \\
 - \left\{ {mc^2\cosh \theta + \hbar c\left[ {\sinh \theta \left( {\frac{d}{dr} + \frac{1}{r}} \right) + \frac{Z\alpha }{r}} \right]} \right\}Q = EQ \\
 \end{array}
\end{equation}

\noindent By using the definition (\ref{eq4}), it becomes
\begin{equation}
\label{eq34}
\begin{array}{l}
 \frac{2ab\hbar c}{a^2 - b^2}\frac{dR}{dr} + \left[ {\frac{\left( {a^2 + b^2} \right)mc^2}{a^2 - b^2} + \left( {\frac{2ab}{a^2 - b^2} - Z\alpha } \right)\frac{\hbar c}{r} - E} \right]R \\
 - \frac{\left( {a^2 + b^2} \right)\hbar c}{a^2 - b^2}\frac{dQ}{dr} - \left[ {\frac{2abmc^2}{a^2 - b^2} + \left( {\frac{a^2 + b^2}{a^2 - b^2} - k} \right)\frac{\hbar c}{r}} \right]Q = 0 \\
 \frac{\left( {a^2 + b^2} \right)\hbar c}{a^2 - b^2}\frac{dR}{dr} + \left[ {\frac{2abmc^2}{a^2 - b^2} + \left( {\frac{a^2 + b^2}{a^2 - b^2} + k} \right)\frac{\hbar c}{r}} \right]R \\
 - \frac{2ab\hbar c}{a^2 - b^2}\frac{dQ}{dr} - \left[ {\frac{\left( {a^2 + b^2} \right)mc^2}{a^2 - b^2} + \left( {\frac{2ab}{a^2 - b^2} - Z\alpha } \right)\frac{\hbar c}{r} - E} \right]Q = 0 \\
 \end{array}
\end{equation}

\noindent Because it is not known for value of the parameter $a$ and
$b$, we cannot obtain the formula of the energy levels in Coulomb
field. According to the conflicting expressions (\ref{eq23}), by
using $a^2 - b^2 = 1$, it can only yield
\begin{equation}
\label{eq35}
\begin{array}{l}
 2ab\hbar c\frac{dR}{dr} + \left[ {\left( {a^2 + b^2} \right)mc^2 - E + \left( {2ab - Z\alpha } \right)\frac{\hbar c}{r}} \right]R \\
 - \left( {a^2 + b^2} \right)\hbar c\frac{dQ}{dr} - \left[ {2abmc^2 + \left( {a^2 + b^2 - k} \right)\frac{\hbar c}{r}} \right]Q = 0 \\
 \left( {a^2 + b^2} \right)\hbar c\frac{dR}{dr} + \left[ {2abmc^2 + \left( {a^2 + b^2 + k} \right)\frac{\hbar c}{r}} \right]R \\
 - 2ab\hbar c\frac{dQ}{dr} - \left[ {\left( {a^2 + b^2} \right)mc^2 + E + \left( {2ab - Z\alpha } \right)\frac{\hbar c}{r}} \right]Q = 0 \\
 \end{array}
\end{equation}

\noindent This is the original shape of the introduced expression
(\ref{eq1}). It is clear that this system of differential equations
are not equivalent to the original Dirac-Coulomb equations, and one
cannot translate them into the so-called Schr\"{o}dinger-like
equations (\ref{eq8}). Only when one knows the value of the
parameters $a$ and $b$ can translate the system of first-order
differential into the corresponding second-order differential
equations without the undetermined parameters. Consequently, the
second-order equations (\ref{eq8}) are the pseudo equations, and the
formula of the energy levels (\ref{eq9}) is not a necessary
mathematical deduction.

\section{Conclusions}

In the present paper, we have shown that the original paper
published in 1985 used many self-contradictor definition expressions
to finally written the so-called simplified solutions of the
Dirac-Coulomb equation and all of the given results in the original
paper cannot be recovered via the strict mathematical calculating.
Such kind of simplified solutions is the pseudo solution of the
Dirac equation in the Coulomb field. The corresponding second-order
Dirac-Coulomb equations are the pseudo second-order Dirac equation.
In fact, it is very simple to solve the system of first-order
Dirac-Coulomb equation with the rough boundary condition or the
exact boundary condition. We have not understood why many papers
treating of the relativistic quantum mechanics seek the
corresponding second-order Dirac equation for writing the so-called
simplified solutions. It should be pointed out that the solutions of
the original system of first-order Dirac-Coulomb equation are not
simplified, and one cannot obtain any simplified solution by
constructing the corresponding second-order Driac-Coulomb equation,
unless introducing some incorrect equations. We can use some
mathematical theorems of the optimum differential
equations\cite{Chen:2000}\cite{Chen:2003} to discuss such kind of
problems\cite{Chen:2001}, and find many papers for constructing
second-order Dirac-Coulomb equations are incorrect in mathematical
and physical signification. It must be ingeminated that the boundary
condition and the uniqueness of solution are very important in
solving wave equation. Some
classical\cite{Dirac:1958}\cite{Roman:1965}\cite{Bjorken:1964} and
modern\cite{Greiner:2000}\cite{Heinrich:2006}\cite{Thaller:1992}
quantum mechanics textbooks have treated these
problems\cite{Gordon:2006}. Not anyone should make any mathematical
mistakes to spell backward the formula of the energy eigenvalues.
The omnifarious mathematical mistakes concealed in the mentioned
paper are cross-sectional, implying there are too much similar
problems in many published
papers\cite{Nakatsuji:2005}\cite{Poszwa:2007}\cite{Pudlak:2006}.


\begin{thebibliography}{10}
\bibitem{Bialynicki:2004} I. Bialynicki-Birula, Particle Beams Guided by Electromagnetic Vortices:New Solutions of the Lorentz, Schrödinger, Klein-Gordon, and Dirac Equations, \textit{Phys.Rev. Lett}. \textbf{93}, 020402 (2004).
\bibitem{Bialynicki:1997} I. Bialynicki-Birula, Rotational frequency shift (with Z.Bialynicka-Birula), \textit{Phys.Rev. Lett}. \textbf{78}, 2539 (1997).
\bibitem{Cooper:2003} F. Cooper, P. Sodano, A. Trombettoni, A. Chodos, An O(N) symmetric extension of the Sine-Gordon Equation, \textit{Phys.Rev. D}, \textbf{68} 045011 (2003)
\bibitem{Darwin:1928} C. G. Darwin, The Wave Equations of the Electron, \textit{Proc. R. Soc. London, Ser. A}, \textbf{118}, 654(1928).
\bibitem{Gordon:1928} W. Gordon, Über den Stoß zweier Punktladungen nach der Wellenmechanik, \textit{Z. Phys}, \textbf{48}, 11(1928).
\bibitem{Chen:2007} R. Chen, New exact solution of Dirac-Coulomb equation with exact boundary condition, \textit{Int. J. Theor. Phys}. (2007-7-30 accepted), arxiv.org/abs/0705.3876.
\bibitem{Chen:1} R. Chen, Established pseudo solution of second-order Dirac-Coulomb equation with position-dependent mass, arxiv.org/abs/0706.4147.
\bibitem{Chen:2} R. Chen, Unheeded pseudo solution of Dirac-Coulomb equations with an indirect transformation of functions, arxiv.org/abs/0707.0091.
\bibitem{Simplified:1985} J. Y. Su, Simplified solution of the Dirac-Coulomb equation, \textit{Phys. Rev. A}, \textbf{32, }3251(1985).
\bibitem{Ciftci:2005} H. Ciftci, R. L. Hall, N. Saad, Iterative solutions to the Dirac equation, \textit{Phys. Rev. A}, \textbf{72}, 022101 (2005).
\bibitem{Boulanger:2006} N. Boulanger, P. Spindel, F. Buisseret, Bound states of Dirac particles in gravitational fields,\textit{ Phys. Rev. D}, \textbf{74}, 125014
(2006).
\bibitem{Alhaidari:2007} A. D. Alhaidari, H. Bahlouli, A. Al-Hasan, and M. S. Abdelmonem, Phys. Rev. A \textbf{75}, 062711 (2007).
\bibitem{Chen:2000} R. Chen, The optimum differential equation, \textit{Chin. J. Engin. Math}. \textbf{17}, 82 (2000) (in Chinese).
\bibitem{Chen:2003} R. Chen, The uniqueness of the eigenvalue assemblage for optimum differential equations, \textit{Chin. J. Engin. Math}. \textbf{20}, 121(2003) (in Chinese).
\bibitem{Chen:2001} R. Chen, The problem of initial value for the plane transverse electromagnetic mode, \textit{Acta Physica Sinica}, \textbf{49} (2000) 2514-2518 (In Chinese).
\bibitem{Dirac:1958} P. A. M. Dirac, \textit{The principles of Quantum Mechanics}, Clarendon Press, Oxford, 1958, pp270.
\bibitem{Roman:1965} P. Roman,\textit{ Advanced Quantum Theory}, Addison-Wesley, 1965.
\bibitem{Bjorken:1964} J. D. Bjorken, S. D. Drell, \textit{Relativistic Quantum Mechanics}, McGraw--Hill, New York, 1964.
\bibitem{Greiner:2000} W. Greiner, \textit{Relativistic Quantum Mechanics: Wave Equations}, Springer, 2000, 3rd Edition, pp225.
\bibitem{Heinrich:2006} Heinrich Saller, \textit{Operational Quantum Theory ¢ò: Relativistic Structures}, Springer, New York, 2006.
\bibitem{Thaller:1992} B. Thaller, \textit{The Dirac Equation}, Springer, New York, 1992.
\bibitem{Gordon:2006} Gordon W. F. Drake, \textit{Handbook of Atomic, Molecular, and Optical Physics}, Springer, New York, 2nd ed., 2006.
\bibitem{Nakatsuji:2005} H. Nakatsuji, H. Nakashima, Analytically Solving the Relativistic Dirac-Coulomb Equation for Atoms and Molecules, \textit{Phys. Rev. Lett}. \textbf{95}, 050407 (2005).
\bibitem{Poszwa:2007} A. Poszwa and A. Rutkowski, Static dipole magnetic susceptibilities of relativistic hydrogenlike atoms: A semianalytical approach, \textit{Phys. Rev. A}, \textbf{75}, 033402 (2007).
\bibitem{Pudlak:2006} M. Pudlak, R. Pincak, V. A. Osipov, Low-energy electronic states in spheroidal fullerenes, \textit{Phys. Rev. B}, \textbf{74}, 235435 (2006)
\end{thebibliography}
\end{document}